\documentclass[11pt]{iopart}
\usepackage{comment}
\usepackage{enumerate}
\usepackage{amssymb}
\expandafter\let\csname equation*\endcsname\relax
\expandafter\let\csname endequation*\endcsname\relax
\usepackage{amsmath}
\usepackage{braket}
\usepackage{dsfont}
\usepackage{graphicx}
\usepackage[usenames,dvipsnames]{color}
\usepackage{tensor}
\usepackage{empheq}
\usepackage{capt-of}
\usepackage[normalem]{ulem} % either use this (simple) or
\usepackage{soul} %
\usepackage{MnSymbol}
\usepackage{cancel}
\usepackage{tikz}

\usepackage[colorlinks,bookmarks=false,citecolor=NavyBlue,linkcolor=OliveGreen,urlcolor=blue]{hyperref}
\newcommand{\be}{\begin{equation}}
\newcommand{\ee}{\end{equation}}
\newcommand{\ba}{\begin{aligned}}
\newcommand{\ea}{\end{aligned}}
\newcommand{\bw}{\begin{widetext}}
\newcommand{\ew}{\end{widetext}}

\newcommand{\bea}{\begin{eqnarray}}
\newcommand{\eea}{\end{eqnarray}}

\newcommand{\II}{\mathrm{i}}
\newcommand{\D}{\mathrm{d}}

\def\doi{http://dx.doi.org/}

\begin{document}
\title{Entanglement in the Quantum Spherical Model - a Review}
\author{Sascha Wald$^{1,2}$, Raul Arias$^3$, and Vincenzo Alba$^4$}
\address{$^1$Statistical Physics Group, Centre for Fluid and Complex Systems,
Coventry University, Coventry, England}
\address{$^2$\ $\mathbb{L}^4$ Collaboration \& Doctoral College for the
Statistical Physics of Complex Systems,
Leipzig-Lorraine-Lviv-Coventry, Europe}
\address{$^3$Instituto de F\'isica La Plata - CONICET and Departamento de
F\'isica, Universidad Nacional de La Plata C.C. 67, 1900, La Plata,
Argentina}
\address{
$^4$Dipartimento di Fisica dell' Universit\`a di Pisa and INFN, Sezione di Pisa, I-56127 Pisa, Italy\\
%* vincenzo.alba@unipi.it
}

% \eads{\\
% \mailto{sascha.wald@coventry.ac.uk},\\
% \mailto{rarias@fisica.unlp.edu.ar},\\
% \mailto{vincenzo.alba@unipi.it
% }}

\begin{abstract}
	We review some recent results on entanglement in the Quantum Spherical Model (QSM). 
	The focus lays on the physical results rather than the mathematical details.
	Specifically, we study several entanglement-related quantities, such as entanglement 
	entropies, and logarithmic negativity, in the presence of quantum and classical 
	critical points, and in magnetically ordered phases. We consider both the short as 
 	well as the long-range QSM. The study of entanglement properties of the QSM is feasible 
	because the model is mappable to a Gaussian system in any dimension. Despite this 
	fact the QSM is an ideal theoretical laboratory to investigate a wide variety of 
	physical scenarios, such as non mean field criticality, the effect of long-range 
	interactions, the interplay between finite-temperature fluctuations and genuine 
	quantum ones. 
\end{abstract}


\maketitle

\clearpage
\setcounter{page}{1}

%##########################
\section{Introduction}

Quantifying entanglement in strongly interacting many-body systems has become an important 
research theme in recent years, and has provided useful insights to understand 
the structure of quantum correlations~\cite{amico-2008,area,calabrese-2009,laflorencie-2016,Horo09}. 
In principle, quantum states are entangled whenever they cannot be written as a product state.
A plethora of entanglement witnesses has been introduced 
to quantify the extent to which quantum states are entangled. 
Widely used tools  that play an important role in entanglement studies comprise, amongst others: R\'enyi entropies, the mutual information, the entanglement negativity, and the entanglement spectrum.
We shall introduce these quantities in Sec.~\ref{sec:qi} in more detail. It is important to note 
that there is not a single entanglement quantifier that works for all setups and systems, reflecting 
the intricacy of quantum entanglement in many-body systems.

Crucially, the study of entanglement in many-body systems
heavily relies on numerical simulations that are quite demanding, even with modern computing hardware.
Thus, obtaining reliable scaling predictions or extracting qualitative behaviors  
in the thermodynamic limit
is challenging and often not attainable. Similar to the study of continuous phase transitions, 
a viable option to overcome computational limitations is to study simplified systems that allow for 
analytical investigations and predictions~\cite{Bax16}. The spherical 
model~\cite{Berl52,Lew52,Ober72,Henk84} has firmly established itself as a reference system whenever
investigations in generic many-body systems prove to be challenging. The spherical model and its quantum
formulation~\cite{Ober72,Henk84,Vojta96,Bien12,Wald15} are analytically solvable in a variety of scenarios, including arbitrary spatial dimension,
temperature and external fields. Moreover, the model possesses a phase transition separating 
a paramagnetic from a ferromagnetic phase that is generally not in the mean-field universality class.

Not surprisingly, the QSM, and closely-related models, proved to be useful to understand 
entanglement properties of quantum many-body systems~\cite{Lu19,Wald20,Wald20-1,alba2020entanglement,Wald23,met09,metlitski-2011,whit-2016}. Crucially, the QSM allows to 
derive the precise finite-size scaling of entanglement-related quantities, often analytically. 
This happens because the QSM is mappable to a Gaussian bosonic system with a constraint. 
This implies that  entanglement 
properties are obtained from the two point correlation functions~\cite{viktor}, 
which are accessible analytically~\cite{Vojta96,Wald15}. 

The aim of this review is to give a few examples of the wide variety of physical scenarios 
where the behavior of entanglement can be addressed in the controlled setup of the QSM, yet 
retaining the complexity of a quantum many-body system. 
Specifically, here we focus on the main results of 
Refs.~\cite{Wald20,Wald20-1,alba2020entanglement,Wald23}.

%Despite the QSM allowing for analytical investigations this does not mean that these are always 
%straightforward. Rather, a heavy machinery of analytical methods concerning asymptotic analysis of sums 
%and integrals is required in order to extract viable information from the model. This may impede the 
%accessible of the results to a larger audience as they usually are overshadowed by their derivations.
%Hence, in this mini-review we ignore the rigorous mathematical foundations on which the results rest 
%and rather highlight our physical findings.

This review is organized as follows. 
In Sec.~\ref{sec:sm} we review relevant properties of the
QSM. In particular, we highlight how the spherical model was conceived, 
how a quantum version is formulated, and sketch how to generally
solve the model. Phase diagrams of all scenarios considered in this review are also discussed.
In Sec.~\ref{sec:qi} we introduce all relevant entanglement-related quantities that we investigate
in this work, such as the entanglement entropy, the negativity, the entanglement spectrum. We 
also briefly discuss how entanglement quantities are calculated in the QSM. 
%a Gaussian system and how the breaking 
%of a continuous symmetry simplifies certain investigations of the entanglement spectrum.
% 
Sections~\ref{sec:3d},~\ref{sec:2d}, \ref{sec:order} and~\ref{sec:1d} are dedicated to the main results.
In Sec.~\ref{sec:3d} we focus on the interplay between quantum 
and classical fluctuations at finite temperature criticality in the three-dimensional  
QSM. In particular, we show that the logarithmic negativity is able to distinguish 
genuine quantum correlations from classical ones. 
In Sec.~\ref{sec:2d} we explore the entanglement gap (or Schmidt gap), which 
is the lowest laying gap of the so-called entanglement spectrum (ES). 
We consider the zero-temperature QSM in two dimensions. By using dimensional 
reduction, we compute the scaling behavior of the entanglement gap at criticality. 
In Sec.~\ref{sec:order} we show that in the ferromagnetic phase the entanglement gap can be written 
in terms of standard magnetic correlation functions, due to the presence of 
a Goldstone mode. 
Finally, in Sec.~\ref{sec:1d} we study the entanglement gap in the one-dimensional QSM 
with long-range interactions and at zero temperature.
In Sec.~\ref{sec:conclusion} we summarize and conclude our results.

% \newpage   

%##################################
\section{Quantum spherical model}
\label{sec:sm}
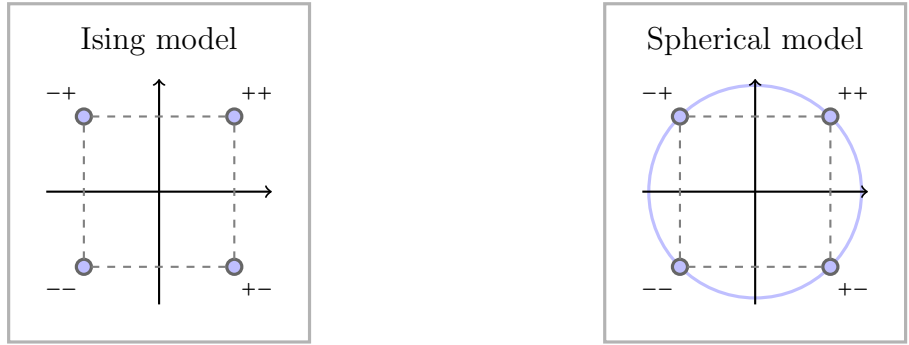
\begin{figure}[t]
\centering
\begin{minipage}{.47\textwidth}
\begin{center}
\begin{tikzpicture}
\filldraw[color=gray!60, fill= white, very thick] (-2,-2) rectangle (2,2.5);
\filldraw[color=black!60, fill=blue!25, very thick](-1,-1) circle (.1);
\filldraw[color=black!60, fill=blue!25, very thick](-1,1) circle (.1);
\filldraw[color=black!60, fill=blue!25, very thick](1,1) circle (.1);
\filldraw[color=black!60, fill=blue!25, very thick](1,-1) circle (.1);
\draw[black, thick,->] (0,-1.5) -- (0,1.5);
\draw[black, thick,->] (-1.5,0) -- (1.5,0);
\draw[gray, thick,dashed](1,-0.9) -- (1,0.9);
\draw[gray, thick,dashed](-1,-0.9) -- (-1,0.9);
\draw[gray, thick,dashed](-0.9,1) -- (0.9,1);
\draw[gray, thick,dashed](-0.9,-1) -- (0.9,-1);
\node[] at (1.3,1.3) {\footnotesize $++$};
\node[] at (-1.3,-1.3) {\footnotesize $--$};
\node[] at (-1.3,1.3) {\footnotesize $-+$};
\node[] at (1.3,-1.3) {\footnotesize $+-$};
\node[] at (0,2) {Ising model};
\end{tikzpicture}
\end{center}
\end{minipage}
\quad
% configuration space Spherical model
\begin{minipage}{.47\textwidth}
\begin{center}
\begin{tikzpicture}
\filldraw[color=gray!60, fill=white, very thick] (-2,-2) rectangle (2,2.5);
\draw[color=blue!25, very thick](0,0) circle (1.4142);
\filldraw[color=black!60, fill=blue!25, very thick](-1,-1) circle (.1);
\filldraw[color=black!60, fill=blue!25, very thick](-1,1) circle (.1);
\filldraw[color=black!60, fill=blue!25, very thick](1,1) circle (.1);
\filldraw[color=black!60, fill=blue!25, very thick](1,-1) circle (.1);
\draw[black, thick,->] (0,-1.5) -- (0,1.5);
\draw[black, thick,->] (-1.5,0) -- (1.5,0);
\draw[gray, thick,dashed](1,-0.9) -- (1,0.9);
\draw[gray, thick,dashed](-1,-0.9) -- (-1,0.9);
\draw[gray, thick,dashed](-0.9,1) -- (0.9,1);
\draw[gray, thick,dashed](-0.9,-1) -- (0.9,-1);
\node[] at (1.3,1.3) {\footnotesize $++$};
\node[] at (-1.3,-1.3) {\footnotesize $--$};
\node[] at (-1.3,1.3) {\footnotesize $-+$};
\node[] at (1.3,-1.3) {\footnotesize $+-$};
\node[] at (0,2) {Spherical model};
\end{tikzpicture}
\end{center}
\end{minipage}

\caption{{\bf Illustration of configuration spaces}:
On the left we show all possible configurations of two Ising spins (vertices of a square). 
On the right we show the extension of the configuration space to two spherical spins.
}
\label{fig:configs}

\end{figure}

The Ising model, see Ref.~\cite{Ising2017} for an overview,
has significantly contributed to our modern understanding of collective phenomena~\cite{Bax16}. 
Despite its apparent simplicity it finds applications in a wide variety of 
fields, see, e.g.,~\cite{Taroni2015,Okamoto2021,Bru67,Bart19,Bot20} for a brief list 
that is by no means exhaustive. 

The classical Ising model is analytically solvable in one spatial dimension, see, 
e.g.,~\cite{Yeo92}, but it does not possess a phase transition. In two dimensions 
the model is still exactly solvable and it exhibits a  finite temperature transition. 
The Ising universality class of the transition is one of the most studied 
in statistical physics~\cite{pelissetto2002critical}. 
Alas, already the three dimensional Ising model is not solved analytically to date~\cite{Nishi10}. 
%This is 
%frustrating since the spatial dimension is one of the few properties that actually affects 
%the universal properties of continuous phase transitions~\cite{Nishi10}. 

To overcome this problem, Berlin and Kac in 1952 suggested a 
generalization of the Ising model by replacing the discrete Ising spin degrees 
of freedom with continuous ones with an additional 
constraint that enforces some of the properties of the original Ising degrees of 
freedom~\cite{Berl52}.
Specifically, since each Ising spin on a lattice $\mathcal{L}$ satisfies $\sigma_i^2=1$,
it is obvious that $\sum_{i\in \mathcal{L}} \sigma_i^2 = V$ with $V$ being the system volume.
Replacing the Ising spins with continuous degrees of freedom, i.e.,
$\sigma_i \to S_i \in \mathbb{R}$, while simultaneously enforcing the external constraint 
$\sum_{i\in \mathcal{L}} S_i^2 = V$ yields the original formulation of the classical spherical model, see Fig.~\ref{fig:configs}.
This classical spherical model is exactly solvable in any  dimension
and supports a finite temperature phase transition in more than two dimensions $d>2$. 
After the paper of Berlin and Kac, later in the same year it was shown that
the strict spherical constraint can be relaxed to be only satisfied on average, i.e.,
$\sum_{i\in \mathcal{L}} \left<S_i^2\right> = V$,
without affecting the universal bulk behavior of the model~\cite{Lew52}. Interestingly, the   
spherical model is related to more realistic spin systems like the $O(N)$ Heisenberg model in the 
limit $N\to\infty$ of infinite spin dimensionality~\cite{Stan68}. 

The spherical model with the average constraint admits also a quantum generalization~\cite{Ober72,Henk84,Vojta96,Bien12,Wald15}. The Hamiltonian of this quantum spherical model (QSM) reads
\begin{align}
	\label{eq:ham}
	H = \sum_{n\in \mathcal{L}}\left( \frac{g}{2} p_n^2 
	+ \frac{1}{2}\sum_{m\in \mathcal{L}}u_{nm} x_n x_m\right),
	\quad \text{with} \quad \sum_{n\in \mathcal{L}} \left< x_n^2\right> = V.
\end{align}
Here the classical spin degrees of freedom $S_i$ are replaced by position operators $x_i$, and  
associate momentum operators $p_i$ were introduced which satisfy $[x_n,p_m] = \II \hbar \delta_{nm}$.
We consider $\mathcal{L}$ as a $d$ dimensional hypercubic lattice. 
\begin{figure}[t]
 \centering
 \includegraphics[width=.32\textwidth]{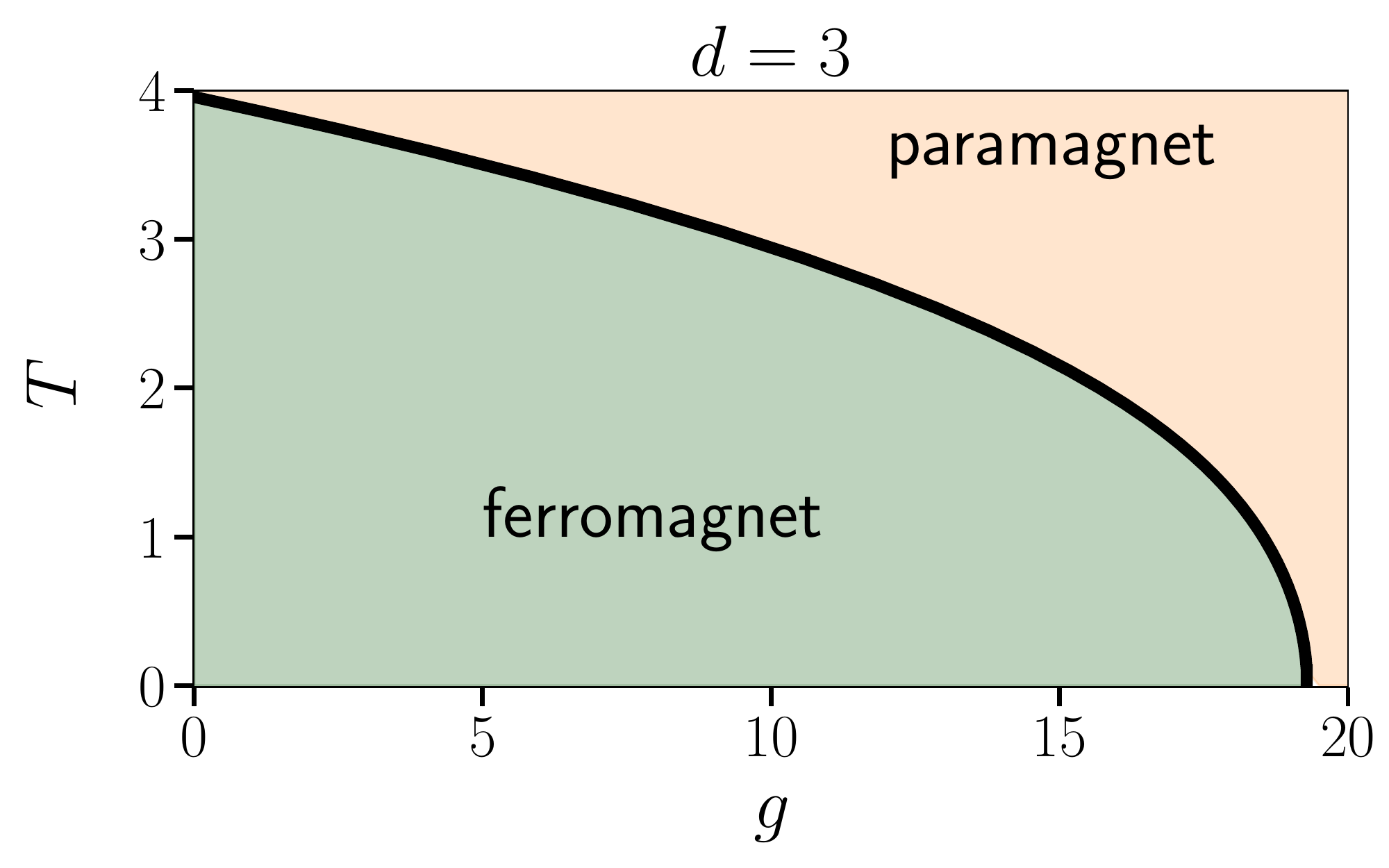}
 \includegraphics[width=.32\textwidth]{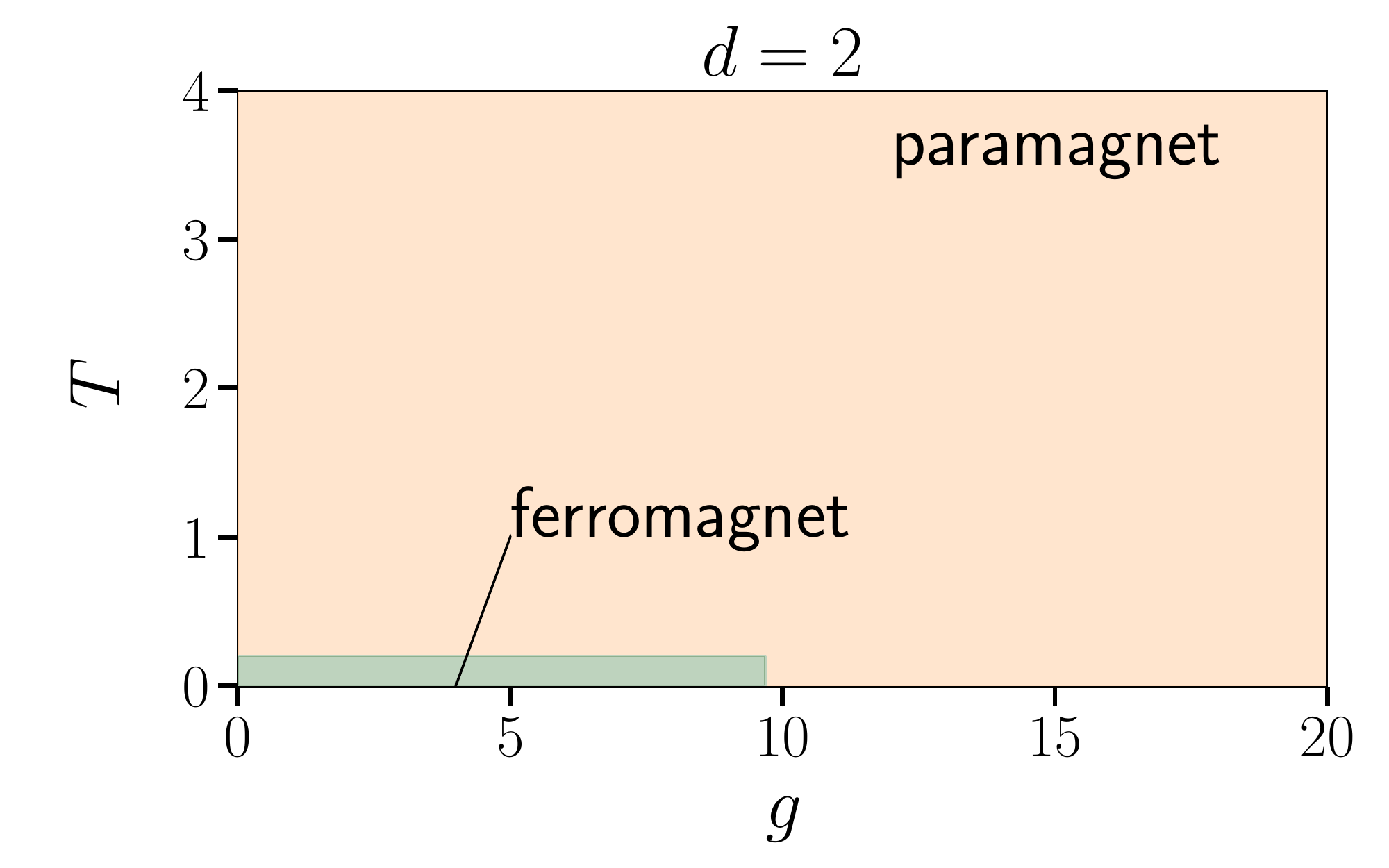}
 \includegraphics[width=.32\textwidth]{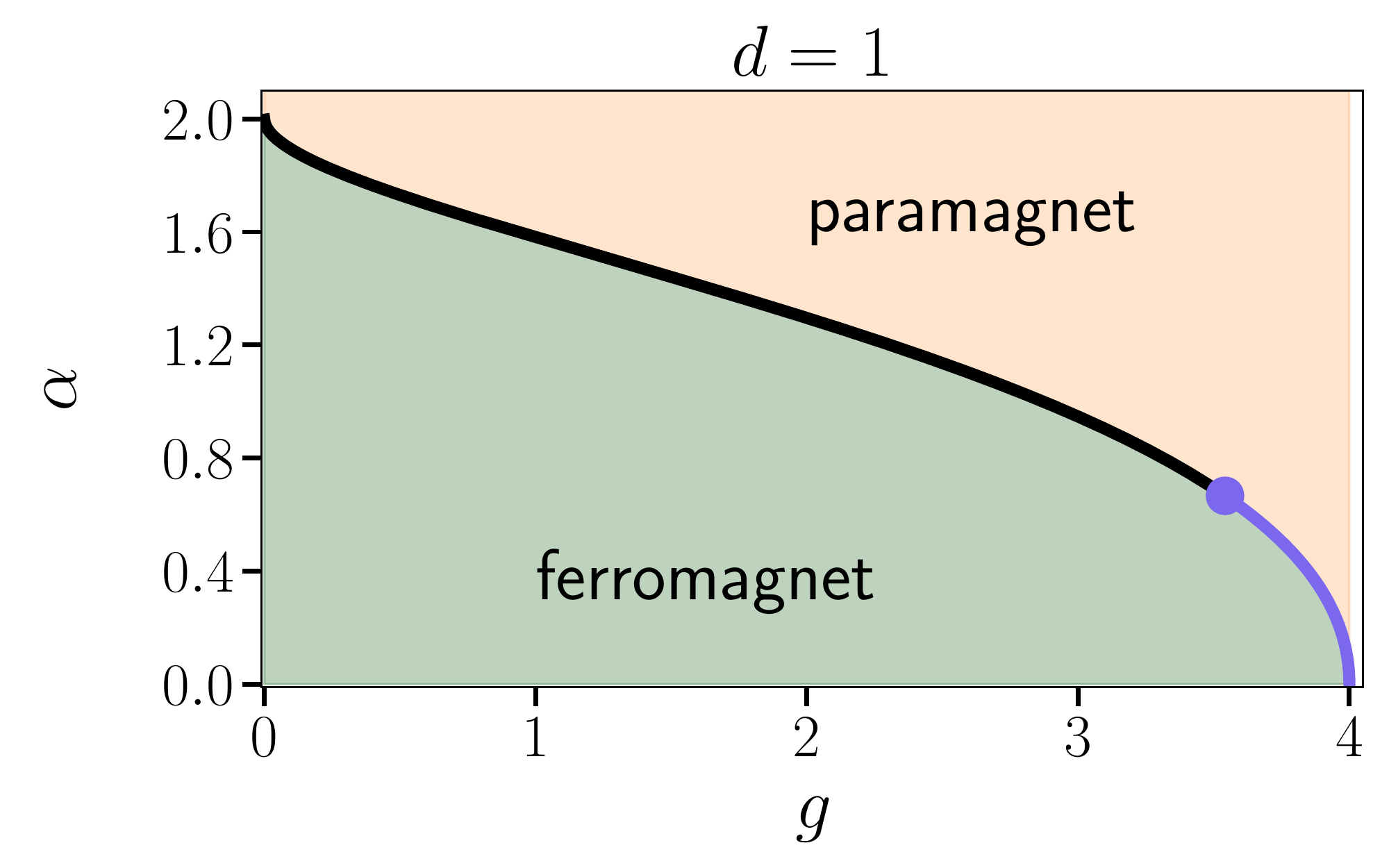}
 \caption{{\bf Phase diagrams of the QSM in different dimension}: In $d=3$ (left panel) 
 the QSM with  nearest neighbor interactions exhibits a thermal critical line and a 
 quantum phase transition at $T=0$. In $d=2$ (center panel)
 only a quantum phase transition is present. For long-range interactions in $d=1$ 
 (right panel) the   
 zero-temperature critical behavior depends on the long-range exponent $\alpha$.  
 For $\alpha<2/3$ the transition is mean-field (purple).
 }
 \label{fig:pds}
\end{figure}

In addition to the finite temperature transitions of the classical spherical model, 
the QSM supports a zero-temperature quantum phase transition~\cite{Vojta96,Wald15}.
If $u_{nm}$ is short-ranged, e.g., nearest neighbor interaction, then the quantum phase 
transition exists for $d>1$ only. This transition is generally in the same universality
class as the thermal transition in the $d+1$ dimensional classical spherical model~\cite{Vojta96}. 
Conversely, for long-ranged interactions a quantum phase transition
is also present for $d=1$~\cite{Vojta96}. The universality class depends on the long 
range exponent $\alpha$ (see below), although in a simple manner. 
Generally, $u_{nn} = 2\mu$ where $\mu$ is a Lagrange multiplier (chemical potential)
that allows to enforce  the spherical constraint. This $\mu$ plays 
the physical role of a mass for the model, or, equivalently, of the inverse correlation 
length. Hence, the critical line of the model is retrieved from the constraint 
for $\mu =0$~\cite{Vojta96,Wald15}.

Here, we only consider translation invariant systems with periodic boundary conditions such 
that the Hamiltonian generally decouples in Fourier space. Let us introduce the Fourier 
transformed operators as 
\begin{align}
 x_n = \frac{1}{\sqrt{V}}\sum_{k\in\mathcal{B}}e^{\II k n} q_k,\qquad 
 p_n = \frac{1}{\sqrt{V}}\sum_{k\in\mathcal{B}}e^{-\II k n} \pi_k
\end{align}
with the $d$ dimensional Brioullin zone $\mathcal{B}$. We  recast the  QSM 
Hamiltonian in the form
\begin{align}
	\label{eq:ham-diag}
 H = \sum_{k\in\mathcal{B}}\frac{g}{2} \pi_k \pi_{-k} + \frac{u_k}{2}  q_k q_{-k}
	= \sum_{k\in\mathcal{B}}E_k \left(b_k^\dagger b_k +\frac{1}{2}\right)
\end{align}
with $u_k$ being the Fourier transform of the interaction potential $u_{nm}$, $E_k = \sqrt{g u_k}$
the eigenenergies of the QSM, and $b_k$, $b_k^\dagger$ are adequately chosen bosonic ladder operators.
This allows us to explicitly write the equilibrium correlation functions 
for a system at temperature $T$ as
\begin{align}
\label{eq:xx}
  \mathbb{X}_{nm}&:= \langle x_{n} x_{m} \rangle= \frac{g}{2V}
 \sum_{k} e^{\II (n-m)k}  \frac{1}{E_k}\coth\left(\frac{E_k}{2T}\right), \\
 \label{eq:pp}
\mathbb{P}_{nm}&:=\langle p_{n} p_{m} \rangle= \frac{1/g}{2V}\sum_{k}
e^{-\II (n-m)k}  E_k \coth\left(\frac{E_k}{2T}\right).
\end{align}
In the 
following sections we shall focus on entanglement quantities in the QSM. Specifically we consider the following situations 
\begin{itemize}
 \item QSM at finite temperature $T>0$ for $d=3$ with nearest neighbor interactions, viz.,
 \begin{align}
u(k) = 2\mu + 2(3-\cos k_x-\cos k_y-\cos k_z)
 \end{align}

 \item QSM at $T=0$ for $d=2$ with nearest neighbor interactions, viz.,
 \begin{align}
u(k) = 2\mu + 2(2-\cos k_x-\cos k_y)
 \end{align}
 
 \item QSM at $T=0$ for $d=1$ with long-range interactions~\cite{Zoi07,Nez12}, viz.,
 \begin{align}
	 \label{eq:long-range}
	 u(k) = 2\mu + \left(2(1-\cos k)\right)^{\alpha/2}.
 \end{align}
 
\end{itemize}
In Eq.~\eqref{eq:long-range}, $\alpha$ is the exponent governing the decay with distance of the 
long-range interactions. 
The phase diagrams for these systems are depicted in Fig.~\ref{fig:pds}. An important ingredient for the 
further analysis is to understand the finite-size scaling  of the spherical 
parameter $\mu$. Specifically, we use that 
$\mu\to0$ in the ferromagnetic phase and at criticality for $L\to\infty$. For finite $L$ conversely,
$\mu$ is always finite. A variety of works have considered this scaling in the classical 
spherical model, see, e.g., Refs.~\cite{Barber73,Bran88,Brez82,Singh85}.

%##################################
\section{Entanglement in the quantum spherical model}
\label{sec:qi}
%(2pgs max)

\begin{figure}
\begin{center}
\includegraphics[width=0.75\linewidth]{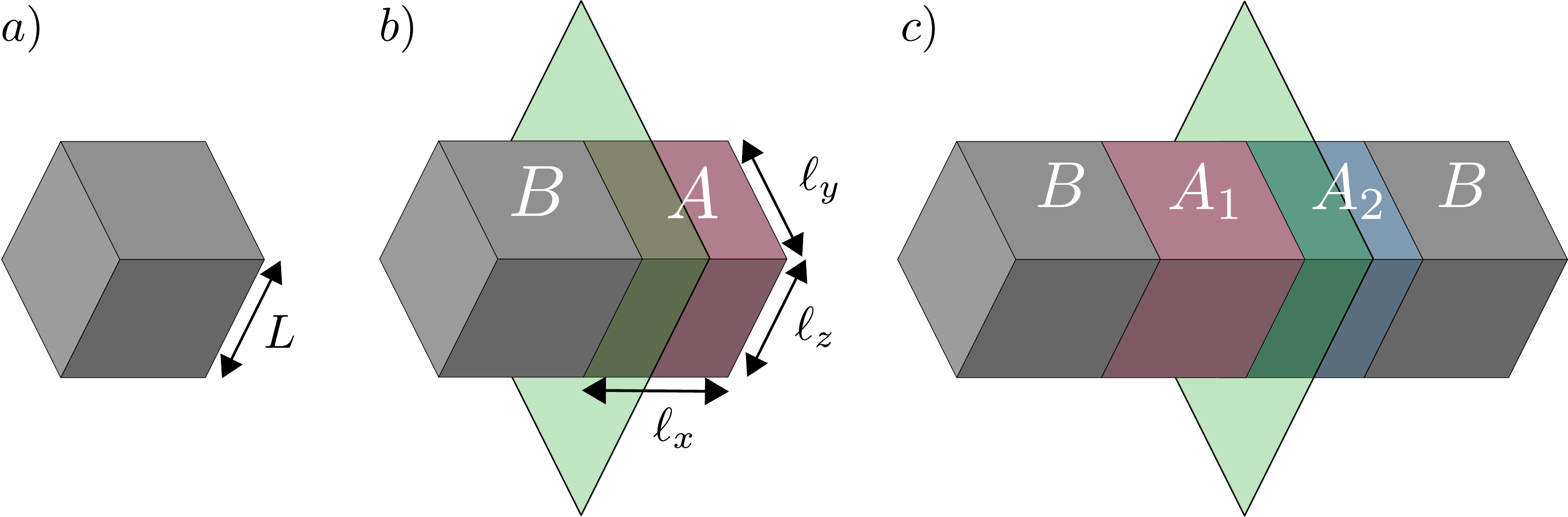}
\caption{{\bf Geometry of partitions}: In panel a) a generic three-dimensional system is 
depicted. In panel b) the system is bipartitioned into two subsystems $A$ and $B$ with $B$ the 
complement of $A$. In panel c) the subsystems $A_1$ and $A_2$ are non-complementary.
}
\label{fig:cube}
\end{center}
\end{figure}

% ----------------------------
In this section we introduce several relevant quantum-information-motivated  
quantities, which  have attracted a lot of
attention in the statistical and high energy theory communities in the last few years. 
Consider a many-body quantum system described by a Hilbert space
${\cal{H}}$ and a density matrix $\rho = \ket{\psi_0}\bra{\psi_0}$ in the 
corresponding zero-temperature
ground state $|\psi_0\rangle$. Upon partitioning the system into two parts $A$ and $B$, see Fig.~\ref{fig:cube}, with 
corresponding Hilbert spaces ${\cal{H}}={\cal{H}}_A\otimes {\cal{H}}_B$ we can define the reduced
density matrix $\rho_A$ of subsystem $A$ by tracing out subsystem $B$, viz.,
\begin{align}
	\rho_A=\mathrm{Tr}_B(\rho).
\end{align}
Although $\rho$ is pure, $\rho_A$ is typically a mixed state because the zero-temperature
ground-state is not separable. In this scenario, the entanglement
entropy
\begin{align}
S_A=-{\mathrm{Tr}}\rho_A\log\rho_A
\end{align}
is a measure of entanglement between the two subsystems. In terms of the 
entanglement spectrum~\cite{laflorencie-2016}, i.e., 
the eigenvalues $\lambda_i$ of the reduced density matrix $\rho_A$, we can
express the entanglement entropy as \cite{ amico-2008, area, laflorencie-2016,Calabrese_2009}
% 
% 
%\begin{align}
%S_A=\sum_j\left[\left(\lambda_j+\frac12\right)\ln\left(\lambda_j+\frac12\right)-\left(\lambda_j-\frac12\right)\ln\left(\lambda_j-\frac12\right)\right].
%\end{align}
% 
% 
% 
% 
\begin{align}
S_A=-\sum_j\lambda_j\ln\lambda_j.
\end{align}
If conversely, the density matrix $\rho$ is not pure, e.g., at finite temperature, 
or if $\rho$ is pure but one is interested in the entanglement between two non-complementary regions  
(see Fig.~\ref{fig:cube} c), then the von Neumann entropy is not a 
good entanglement witness. A useful quantifier in these cases is the
logarithmic negativity~\cite{Lee00,Eis99,Vidal02,Plenio03,Per96,Zyc98}.

The negativity is defined from the so-called partially transposed reduced density 
matrix. Given a partition of $A$ as $A = A_1\cup A_2$ 
(see Fig.~\ref{fig:cube} $c)$), the matrix elements of the partial transpose
$\rho_A^{T_2}$ with respect to the degrees of freedom of $A_2$ are defined as
% 
% 
% Sometimes, the state in the full Hilbert space is not pure, but a mixed state. In such cases the entanglement entropy does not give us any meaningful information 
% about the degree of entanglement between the degrees of freedom in $A$ and those in $B$. In those cases an useful quantity is the logarithmic negativity. It allows, 
% for instance, to quantify the entanglement in a bipartite system at finite temperature, or the entanglement between two non-complementary subsystems at zero temperature. 
% The negativity is defined from the so-called partial transpose. Given the partition of $A$ as $A = A_1\cup A_2$ (see \ref{fig:cube}$(c)$ for an example), the matrix elements
% of the partial transpose $\rho_A^{T_2}$ with respect degrees of freedom of $A_2$ are defined as
% 
% 
\begin{align}
\langle\varphi_1\varphi_2|\rho_A^{T_2}|\varphi_1'\varphi_2'\rangle := \langle\varphi_1\varphi_2'|\rho_A|\varphi_1'\varphi_2\rangle.
\end{align}
Here, $\{\varphi_1\}$ and $\{\varphi_2\}$ are orthonormal bases for $A_1$ and $A_2$ respectively. 
In contrast to the eigenvalues of the reduced density matrix $\rho_A$, the eigenvalues $\zeta_i$ of $\rho_A^{T_2}$ can be positive or negative. The logarithmic negativity is then defined as 
\begin{align}
{\cal{E}}_{A_1:A_2}=\ln\mathrm{Tr}|\rho_A^{T_2}|. 
\end{align}
The behavior of the logarithmic negativity has been fully characterized in systems that are described 
by conformal field theory at zero temperature~\cite{Calabrese2012} and at finite 
temperature~\cite{Calabrese_2015}. Generally, the negativity follows an area law scaling 
as observed in a variety of systems, see, e.g.,
Refs.~\cite{Nobili_2016,Eisler16,Shapourian2019-1,Lu19,Lu20,Wald20}.

Finally, we introduce the entanglement spectrum (ES), viz.,
$\{\xi_i=-\ln(\lambda_i)\ | \ \lambda_i\in \mathrm{spec}(\rho_A)\}$.
The lowest entanglement gap (Schmidt gap) is defined by
\begin{align}
	\label{eq:e-gap}
\delta\xi=\xi_1-\xi_0,
\end{align}
where $\xi_0$ and $\xi_1$ are the lowest and the first excited ES level, respectively.

The ES has received a lot of attention following the observation that it contains universal
information about the edge modes in fractional quantum Hall systems~\cite{li-2008}. Subsequently, 
the ES was investigated in a variety of setups, e.g., in conformal field 
theory~\cite{calabrese-2008,lauchli-2013,Alba-2017,cardy-talk},
in quantum Hall systems~\cite{thomale-2010,andreas-2010,haque-2007,thomale-2010a,hermanns-2011,chandran-2011},
in frustrated and 
magnetically ordered systems~\cite{poilblanc-2010,cirac-2011,de-chiara-2012,alba-2011,metlitski-2011,alba-2012,Alba13,lepori-2013,james-2013,kolley-2013,Chan14,rademaker-2015,kolley-2015,frerot-2016} or 
systems with impurities~\cite{bayat-2014}.

%#############################################################
%#############################################################

The main topic of this review is to investigate the entanglement-related  quantities 
introduced above in the QSM in $d=1$, $d=2$ and $d=3$ spatial dimensions.

Since the QSM is mappable to a Gaussian system of bosons (cf. Eq.~\eqref{eq:ham-diag}), 
entanglement-related quantities can be extracted from the 
position and momentum correlators (cf. Eqs.~\eqref{eq:xx} and~\eqref{eq:pp})  
${\mathbb{X}}\equiv\langle x_n x_m\rangle$
and ${\mathbb{P}}\equiv\langle p_n p_m\rangle$ (see~\cite{viktor} for a review).
First, we consider  the correlators restricted to subsystem $A$, denoting them 
as ${\mathbb{X}}[A]$ and ${\mathbb{P}}[A]$. The single-particle eigenvalues 
$\epsilon_i$, with $i\in[1,|A|]$,  of the entanglement Hamiltonian $H_A$, which is defined as 
$\rho_A=\exp(-H_A)$, are readily related to the eigenvalues $e_i$ of the matrix product
$\mathbb{C}_A = {\mathbb{X}}[A]{\mathbb{P}}[A]$, viz.,
\begin{align}
	\label{eq:eiepsi}
\sqrt{e_i}=\frac12 \coth\left(\frac{\epsilon_i}{2}\right).
\end{align}
The eigenvalues of the entanglement Hamiltonian $H_A$ are constructed by filling 
the single-particle levels $\epsilon_i$ in all possible ways. This allows also to obtain the 
von Neumann entropy $S_{\mathrm{vN}}$ in terms of the eigenvalues $e_j$ as 
\begin{equation}
	S_{\mathrm{vN}}=\sum_j\left[\left(\sqrt{e_j}+\frac{1}{2}\right)
		\ln\left(\sqrt{e_j}+\frac{1}{2}\right)- 
	\left(\sqrt{e_j}-\frac{1}{2}\right)
\ln\left(\sqrt{e_j}-\frac{1}{2}\right)
\right]. 
\end{equation}
Hence, diagonalizing $\mathbb{C}_A$ allows us to deduce the full entanglement spectrum. In 
particular, assuming that the single-particle entanglement spectrum levels are 
ordered as $\epsilon_1\le\epsilon_2\le\cdots\le\epsilon_{|A|}$, 
the Schmidt gap is simply given by $\delta \xi = \epsilon_1$.
For Gaussian bosonic systems, the logarithmic negativity can be constructed from 
the two-point correlation functions~\cite{audenaert2002entanglement}. 
First, we define the transposed matrix ${\mathbb{P}}[A^{T_2}]$ as 
\begin{align}
{\mathbb{P}}[A^{T_2}]\equiv {\mathbb{R}}[A^{T_2}]{\mathbb{P}}[A]{\mathbb{R}}[A^{T_2}],
\end{align}
where the matrix ${\mathbb{R}}[A^{T_2}]$ acts as the identity matrix on $A_1$ and as minus the identity 
matrix on $A_2$. The eigenvalues $\nu_i^2$ of ${\mathbb{X}}[A]{\mathbb{P}}[A^{T_2}]$ form the 
single-particle negativity spectrum. In terms of them the negativity can be written as~\cite{audenaert2002entanglement} 
\begin{align}
{\cal{E}}=\sum_i{\mathrm{max}}(0,-\ln(2\nu_i)). 
\end{align}

\section{Quantum and classical fluctuations at finite temperature criticality}
\label{sec:3d}
\begin{figure}[t]
\begin{center}
 \includegraphics[width=0.9\textwidth]{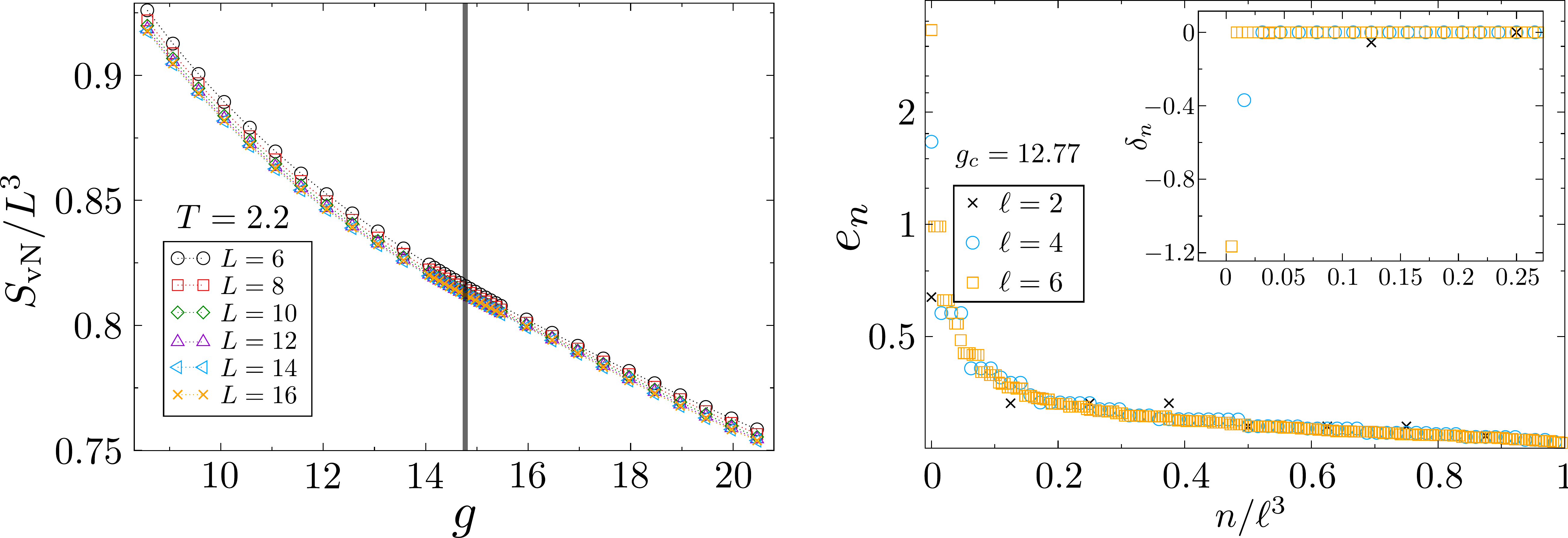}
 \caption{{\bf von Neumann entropy and entanglement spectrum}:
	 (Left panel) Volume-law scaling of the von Neumann entropy across the 
	 finite-temperature transition in the three-dimensional QSM. 
	 Singular terms are present but they vanish at the transition and are 
	 overshadowed by the analytic background. Non-analytic contributions
 	are more visible in the entanglement spectrum (notice the divergence  
       of the largest negative single-particle ES level in the right panel).
       In the inset we show $\delta_n$ (cf. Eq.~\eqref{eq:delta}), which measures 
       the nonanalyticity of the levels across the transition. Here $\delta_n\ne0$ 
       signals nonanalytic behavior.}
\label{fig:S}
\end{center}
\end{figure}
Understanding the interplay of classical and quantum fluctuations is an important but challenging 
task~\cite{Hauke2016,Gabbrielli2018,frerot-2016}. One way of approaching this question is by studying entanglement witnesses in the 
vicinity of a finite temperature phase transition that is driven by classical fluctuations. It 
has been observed that a variety of entanglement witnesses are sensitive to classical criticality. 
For instance, it has been shown that the negativity develops cusp-like 
singularities~\cite{Lu19,Lu20}. In this section we review our investigation
from Ref.~\cite{Wald20} of entanglement-related quantities at the 
finite-temperature transition in the $d=3$ dimensional QSM, see
Fig.~\ref{fig:pds} (left panel).

First, we discuss the von Neumann entropy $S_{\rm vN}$ for the bipartition of the system into 
two equal parts (see Fig.~\ref{fig:cube}). 
As we mentioned in Sec.~\ref{sec:qi}, $S_{\rm vN}$ is not a valid entanglement witness at  
finite temperature. In fact, the von Neumann entropy becomes the standard thermal 
entropy at finite temperature. Indeed, as shown in Fig.~\ref{fig:S} (left panel), 
$S_{\mathrm{vN}}$ satisfies a standard volume-law scaling. Being sensitive to both quantum 
and classical correlations, the von Neumann entropy overestimates the amount of entanglement, 
which is expected to scale with the boundary between the two subsystems. 
Moreover, the von Neumann entropy does not show any singularity at the transition. This 
happens because singular terms, although they are present, vanish at the critical point, and 
are overshadowed by the analytic background. 
Singularities are more visible in the single-particle entanglement spectrum, as 
illustrated in the right panel of Fig.~\ref{fig:S}. In the figure we show  
the entanglement spectrum for two 
adjacent blocks of linear size $\ell$ embedded in an infinite system. The eigenvalues quickly decay upon 
increasing their index and most of them satisfy $e_n\approx 1/4$. 
Clearly, only those eigenvalues
with low index can yield potentially singular contributions. In the inset we investigate the 
singularity using the quantity
\begin{align}
	\label{eq:delta}
 \delta_n = (e_n)_+'-(e_n)_-' 
\end{align}
that measures the difference of the right and left derivatives of $e_n$ 
with respect
to $g$ at $g_c$. Clearly $\delta_n \neq 0$ indicates a non-analyticity and we observe this 
for small $n$.

\begin{figure}[t]
 \centering
 \includegraphics[width = .3\textwidth]{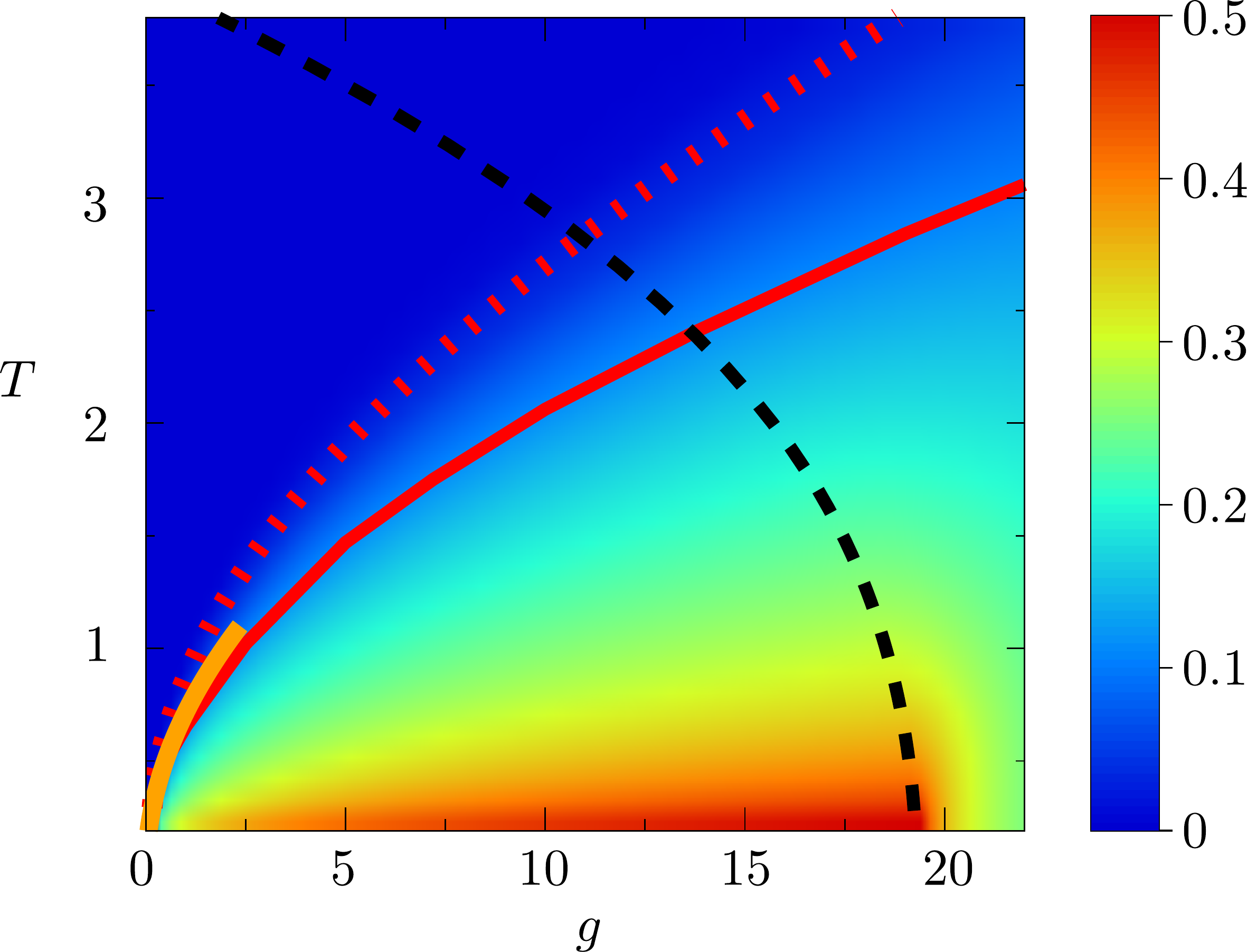}
 \includegraphics[width = .66\textwidth]{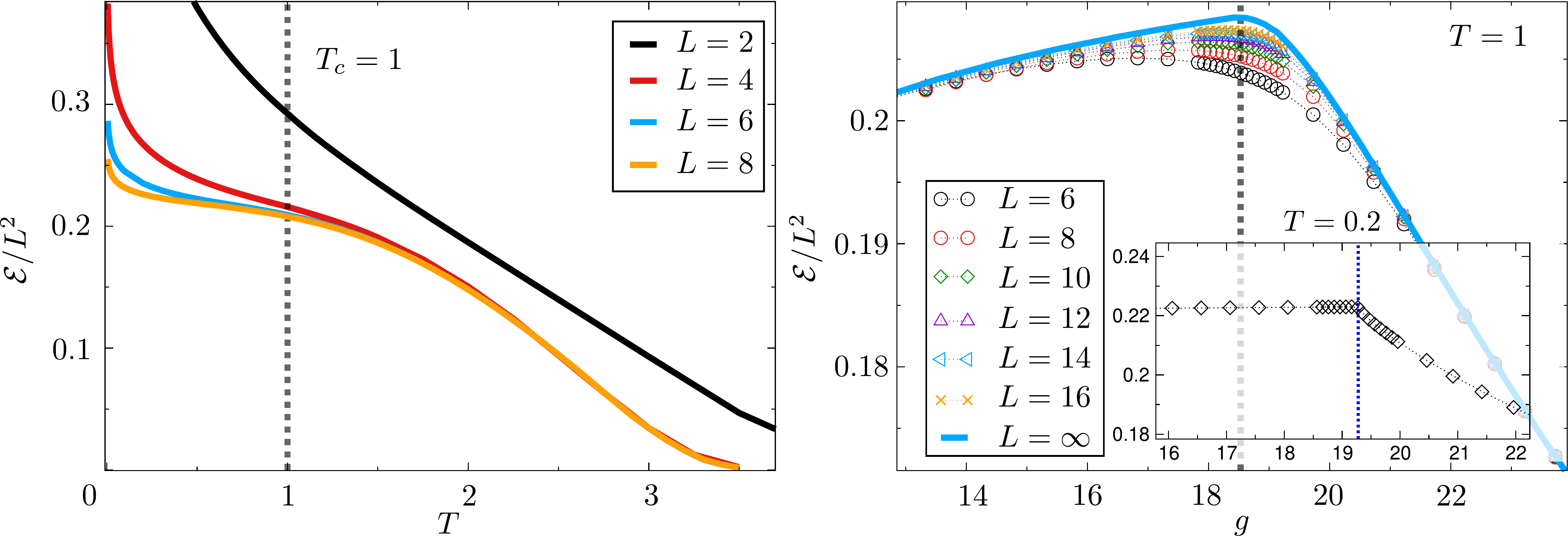}
 \caption{{\bf Entanglement negativity in the three-dimensional QSM}: The left panel is an overview 
	 of the negativity in the $g-T$ plane, with $T$ the temperature and $g$ the quantum coupling (cf. Eq.~\eqref{eq:ham}). Here we always consider the half-system negativity. The critical line is reported as black dashed line.  
	 The red-dashed line is the death line above which the negativity is exactly zero. 
	 The 
	 death line as extracted from the negativity between two adjacent spins is reported as red solid line. The orange solid line is the behavior as $\sqrt{g}$ at small $g$. 
	 The centre (right) panel shows the negativity across the phase transition varying the 
 temperature (quantum parameter). In the right panel the continuous line is the result in 
 the thermodynamic limit. The inset shows the behavior of the negativity at $T=0.2$, i.e., close to  
 the quantum phase transition. 
 }
 \label{fig:negativity}
\end{figure}
Next, we discuss the entanglement negativity. As outlined in Sec.~\ref{sec:qi}, the negativity is a 
proper entanglement witness, and as such obeys an area law, see center and right panel in 
Fig.~\ref{fig:negativity}. Crossing the thermal transition at fixed finite $T$, i.e., changing  
the quantum driving parameter $g$ the negativity decays slowly as $1/g$ for large $g$, 
see Fig.~\ref{fig:negativity} right panel. This is in contrast to the behavior when 
crossing the transition with the temperature $T$ at fixed $g$, as depicted in the center panel of 
Fig.~\ref{fig:negativity}. Here, the negativity shows a sudden death after
and remains exactly zero for increasing $T$. We also see that the negativity does not show any 
cusp singularity across the finite temperature transition but develops a cusp when approaching 
low temperatures (see inset right panel in Fig.~\ref{fig:negativity}). This signals that 
singularities, although present, are strongly suppressed. 
Furthermore, in Fig.~\ref{fig:negativity} (left panel), we map 
out the negativity in the full phase diagram. In the figure the dashed line is the 
critical line separating the paramagnetic and the ferromagnetic phase. 
We observe that the negativity generally attains
a maximum at the quantum phase transition, hinting at a strongly 
entangled quantum state. We also observe 
that the negativity increases upon lowering the temperature and is largest for $T=0$.
We also highlight the numerical death line in Fig.~\ref{fig:negativity} (dotted line in the 
left panel) above which the negativity is zero. 

Interestingly, most of these findings can be quantitatively understood 
considering two adjacent sites embedded in 
an infinite system. This setup allows for analytic investigations as shown in Ref.~\cite{Wald20}.
For large $g$ and constant $T$, this approach qualitatively predicts the slow negativity decay
from Fig.~\ref{fig:negativity} (right panel), viz.,
\begin{align}
 \mathcal{E} = -\ln \left(\frac{g-2}{g}\right) \stackrel{g\to\infty}{\simeq} \frac{2}{g}.
\end{align}
Similarly, it predicts the existence of the death line (continuous line in the figure), 
and correctly captures its onset for small $g$ as $\sqrt{g}$, see 
Fig.~\ref{fig:negativity} (left panel).

\section{Entanglement gap at 2D quantum criticality}
\label{sec:2d}
As we have seen in Sec.~\ref{sec:3d} the low-laying entanglement spectrum encodes relevant 
information about the critical properties of the system. To further investigate this aspect we 
review in this section our studies in Refs.~\cite{Wald20-1,alba2020entanglement} 
of the behavior of the
Schmidt gap at quantum criticality and in the ferromagnetic phase 
in the two dimensional QSM. Interest in the behavior of the Schmidt gap has spiked in the last 
decade. 
%Notice that rks have argued that the information encoded in $\delta\xi$ may in fact
%not be universal~\cite{Chan14}. Hence, relaying on the entanglement gap to detect criticality
%may lead to spurious transitions. 
% 
% 
\begin{figure}[t]
\centering
\includegraphics[width =\textwidth]{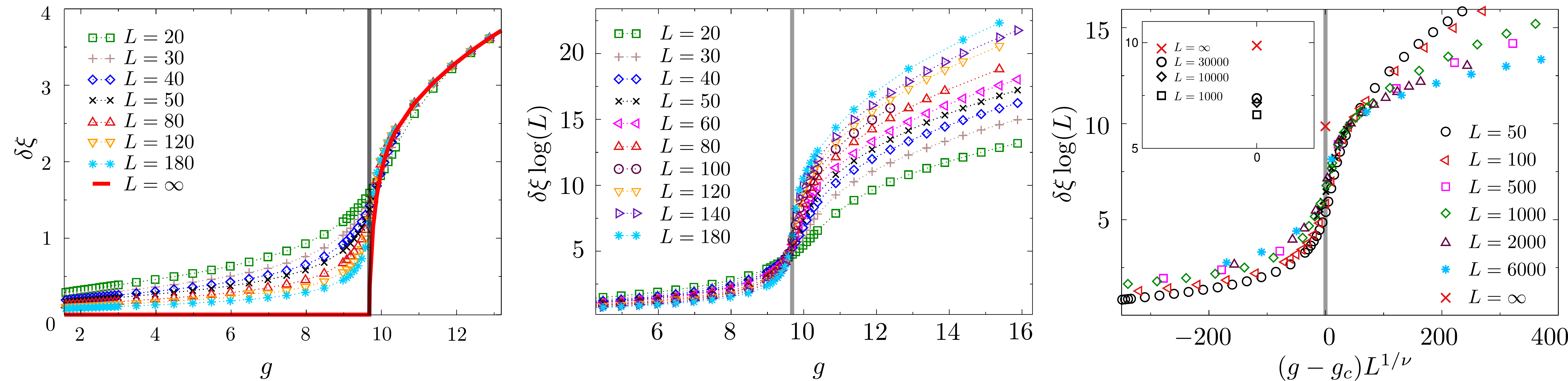}
\caption{{\bf Overview of the entanglement gap in the two-dimensional QSM}:
The left panel shows the gap $\delta\xi$ across the whole phase diagram. We observe distinct behaviors
in the paramagnetic phase (finite gap) and the ferromagnetic phase (vanishing gap). 
We always consider the half-system entanglement spectrum. The continuous line in the left panel 
is the result in the thermodynamic limit~\cite{Wald20-1}. 
%The upmost inset
%illustrates that a crossing of the finite size data is observed in the vicinity of the quantum critical point. The second inset shows the slow vanishing of the gap at the critical point. 
In the center panel we show the rescaled gap $\delta\xi\ln(L)$. At the critical point the gap 
decays as $1/\ln(L)$, which implies that the rescaled gap should exhibit a 
crossing. In the right panel we plot $\delta\xi\ln(L)$ versus $(g-g_c)L^{1/\nu}$. Upon approaching 
the thermodynamic limit $L\to\infty$ a data collapse is expected. The cross symbol is $\delta\xi\ln(L)$ at $g=g_c$ and $L\to\infty$. Still, subleading terms are too
large to observe the collapse, as confirmed in the inset. 
}
\label{fig:2dgap}
\end{figure}

\begin{figure}[t]
 \centering
 \includegraphics[width = .9\textwidth]{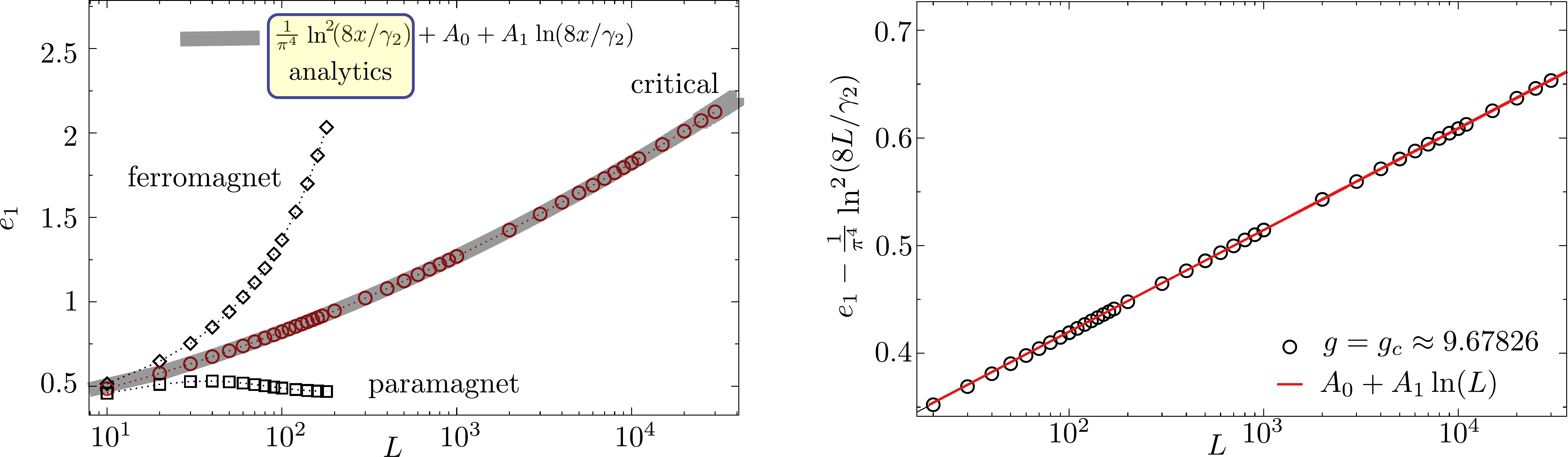}
 \caption{{\bf Scaling of the largest eigenvalue $e_1$ of $\mathbb{C}_A$}: The left panel shows the 
 scaling of $e_1$ in the paramagnetic and the ferromagnetic phase as well as at criticality
 in the $2d$ QSM. In the 
 paramagnetic phase $e_1$ converges to a finite value and in the ferromagnetic phase $e_1$ 
 diverges (hence $\delta\xi\to0$). At criticality we see a slower divergence than in the ordered 
 phase. The full grey line is a fit to $1/\pi^4\ln(8L/\gamma_2)+A_0+A_1\ln(8L/\gamma_2)$, with $A_0,A_1$ 
 fitting parameters. The leading part $\sim\ln^2(L)$ is obtained analytically. Here $\gamma_2$ is a 
 known constant~\cite{Wald20-1}. In the right panel we investigate the subleading logarithmic term.
 }
 \label{fig:2d-ev}
\end{figure}

In the left panel of Fig.~\ref{fig:2dgap} we show the numerical findings for 
the behavior of $\delta\xi$ across the phase diagram, see Fig.~\ref{fig:pds}.
Here, we consider a bipartition
of the system into two equal halves.
In the paramagnetic 
regime, we observe that the gap converges rapidly to a finite number upon 
increasing the linear system size $L$. Hence, the gap remains finite in the thermodynamic
limit $L\to\infty$. Conversely, the behavior at the critical point~\cite{Wald20-1}
and in the ferromagnetic
phase~\cite{alba2020entanglement} differs from that in the paramagnetic phase.
In the ferromagnetic phase, the entanglement gap scales as~\cite{alba2020entanglement}
\begin{align}
	\label{eq:dxi-ferro}
	\delta\xi \simeq \frac{\Omega}{\sqrt{L \ln (L)}}, 
\end{align}
where the constant $\Omega$ is known analytically~\cite{alba2020entanglement}, and 
depends on low-energy properties of the QSM and on the 
geometry.  For instance, $\Omega$ is sensitive to the presence of corners in the boundary 
between $A$ and the rest. 
Hence, the gap closes algebraically, involving logarithmic corrections. At criticality
we find that the Schmidt gap still closes, i.e., $\delta\xi\to0$ albeit significantly slower.
Precisely, the gap vanishes as~\cite{Wald20}
\begin{align}
\label{eq:xi-crit}
 \delta\xi \simeq \frac{\pi^2}{\ln(L)}.
\end{align}
This result is obtained for the bipartition into equal halves as follows. 
Since we use periodic boundary conditions in both directions, and the bipartition does 
not introduce corners, the momentum $k_y$ 
remains a good quantum number also for the reduced density matrix. This  
allows to exploit dimensional reduction~\cite{sara2d} mapping the problem to a one-dimensional 
one.
Hence, we may use the analytical 
result for a one dimensional massive harmonic chain~\cite{viktor}
in order to obtain Eq.~(\ref{eq:xi-crit}).
Since the harmonic chain result is derived using the corner transfer matrix on two infinite halves, 
whereas we have periodic boundary conditions also along the $x$ direction, 
Eq.~(\ref{eq:xi-crit}) is exact only at leading order in $L$. Our results are numerically 
confirmed in Fig.~\ref{fig:2d-ev}. In the figure we consider the largest eigenvalue $e_1$ 
of $\mathbb{C}_A$ (see Section~\ref{sec:qi}).  This is related to the entanglement gap 
via Eq.~\eqref{eq:eiepsi} and Eq.~\eqref{eq:e-gap}. In particular, a diverging $e_1$ implies a 
vanishing entanglement gap. 
Fig.~\ref{fig:2d-ev} shows that the leading behavior of $e_1$ at large $L$ is correctly
captured by the analytic result (full line).  Again, Fig.~\ref{fig:2d-ev} confirms that the 
entanglement gap is finite in the paramagnetic phase, whereas in the ordered phase a faster 
divergence is observed (cf. Eq.~\eqref{eq:dxi-ferro}). 

In the right panel of Fig.~\ref{fig:2d-ev} we subtract the analytic prediction for the 
leading behavior of $e_1$.  The continuous line is a fit to $A_0+A_1\ln(L)$. The result 
of the fit confirms the presence of a logarithmic correction to the leading behavior. 

\section{Entanglement gap and symmetry breaking in the  QSM}
\label{sec:order}

In the ferromagnetically ordered phase of the QSM, the dispersion develops a zero mode, which 
reflects the Goldstone mode associated to symmetry breaking. This implies that 
the position correlation function (see Eq.~\eqref{eq:xx}) diverges. 
Here we show that this  fact is sufficient 
to fully determine the scaling of the entanglement gap. First, the divergence in Eq.~\eqref{eq:xx}
is reflected in the fact that the eigenvector of $\mathbb{C}_A$ associated with the largest 
eigenvalue becomes flat in the thermodynamic limit~\cite{Wald20-1,alba2020entanglement}. 
Hence, we may rewrite the position correlator up to leading order as
\begin{align}
	\mathbb{X} \simeq \chi^x \ket{1}\bra{1},\quad\mathrm{with}\,\,\chi^x\propto L, 
\end{align}
with $\ket{1}$ being a normalized flat vector. First, we note that $\bra{1}\mathbb{X}\ket{1} 
=\chi^x$. Furthermore, it is easy to verify that 
$\bra{1} \mathbb{P}$ and $\ket{1}$ are left and right eigenvectors of $\mathbb{C}$. Both correspond 
to the largest (diverging) eigenvalue $e_1$. Hence, it is straightforward to identify 
\begin{align}\label{eq:e1}
	e_1 = \bra{1}\mathbb{X}\ket{1} \bra{1}\mathbb{P}\ket{1}:=\chi^x\chi^t,  
\end{align}
where $\chi^t:=\langle1|\mathbb{P}|1\rangle$. Here we should observe that $\chi^x$ resembles 
a susceptibility for the position variables $x_i$, whereas $\chi^t$ is the susceptibility 
of the $p_i$. Eq.~\eqref{eq:e1} establishes a remarkable correspondence between the entanglement 
gap and standard quantities such as $\chi^x$ and $\chi^t$. A similar decomposition as in
Eq.~\eqref{eq:e1} 
was employed in Ref.~\cite{Botero04} to treat the zero-mode contribution to entanglement in 
the harmonic chain.

\begin{figure}[t]
 \centering
 \includegraphics[width=.5\textwidth]{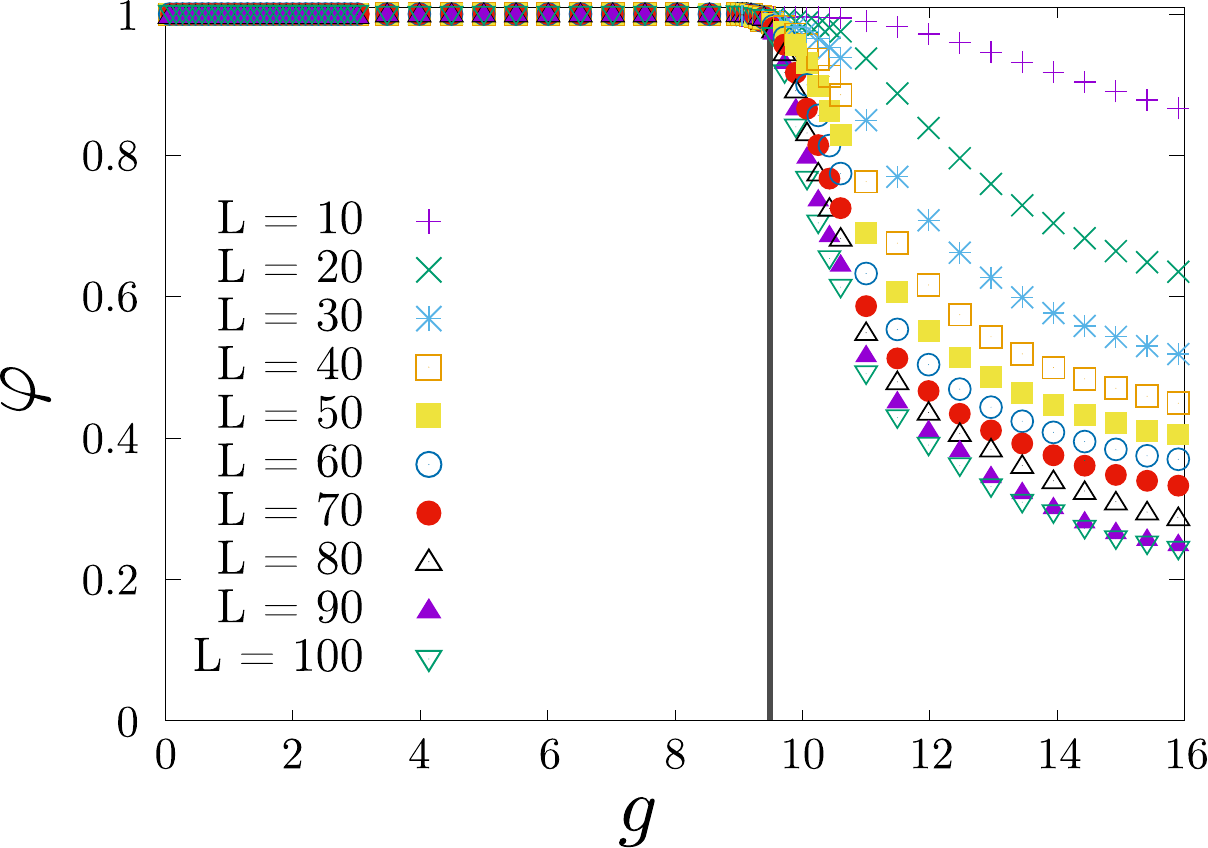}
 \caption{{\bf Lowest eigenvector of $\mathbb{C}_A$}: We show the overlap $\varphi$ 
	 between the eigenvector of $\mathbb{C}_A$ corresponding to the largest eigenvalue and 
	 the flat vector. In the ordered phase of the QSM one has that $\varphi\to1$ in the 
	 thermodynamic limit. This is not the case at the critical point and in the ferromagnetic 
	 phase.
 }
	 \label{fig:flat}
\end{figure}

We now verify numerically that, as expected, the eigenvector of $\mathbb{C}_A$ corresponding 
to its largest eigenvalue $e_1$ 
becomes flat in the ferromagnetic phase, which ensures the validity of Eq.~\eqref{eq:e1}. 
Our results are reported in Fig.~\ref{fig:flat} where we show the overlap 
 $\varphi$ between the flat vector and the exact eigenvector of the 
 correlation matrix. 
 At criticality, the eigenvector is not flat in the thermodynamic limit $L\to\infty$ 
 (see Fig.~\ref{fig:flat}). 
 On the other hand, below the critical point the eigenvector becomes flat in the 
 thermodynamic limit. 

\subsection{Entanglement gap in the ferromagnetic phase of the 2D QSM}

Let us now discuss the application of Eq.~\eqref{eq:e1} in the ordered phase of the two-dimensional 
QSM. 
To employ Eq.~\eqref{eq:e1}  we have  to evaluate the 
flat vector expectation values of the position and momentum correlation matrix.
The standard way is to decompose them in a thermodynamic and a finite size part, i.e., 
\begin{align}
 \bra{1}\mathbb{X}\ket{1} = \bra{1}\mathbb{X}^{\rm (th)}\ket{1} + \bra{1}\mathbb{X}^{(L)}\ket{1}
\end{align}
and similar for $\bra{1}\mathbb{P}\ket{1}$, using the Poisson summation formula 
\begin{equation}
\label{eq:poisson}
 \sum_{n=a}^b f(n) = \frac{f(a)+f(b)}{2} + \int_a^b f(x) {\D}x
 +2\sum_{p=1}^\infty \int_a^b f(x) \cos(2\pi p x ) {\D} x.
\end{equation}
The FSS of these contributions is then obtained from the FSS of the spherical parameter $\mu$ and 
from standard methods such as stationary phase methods, Euler-Maclaurin formulas, and 
Mellin transform techniques. One finds~\cite{alba2020entanglement} 
\begin{align}
 \bra{1}\mathbb{X}^{\rm (th)}\ket{1} \sim L^2, \quad
 \bra{1}\mathbb{X}^{(L)}\ket{1} \sim L^2, \quad
 \bra{1}\mathbb{P}^{\rm (th)}\ket{1} \sim \frac{\ln(L)}{L}, \quad
 \bra{1}\mathbb{P}^{(L)}\ket{1} \sim \frac{\ln(L)}{L}.
\end{align}
Notice that $\chi^t$ vanishes at $L\to\infty$. 
Thus, the entanglement gap scales as $\delta \xi \sim 1/\sqrt{L\ln(L)}$. This is in 
agreement with our numerical simulations~\cite{alba2020entanglement}. Furthermore, 
it has been suggested
that for continuous symmetries, the gap should vanish as~\cite{metlitski-2011} 
$\delta \xi \sim 1/(L\ln(L))$ which differs
from our result. This could be specific of the QSM, although the issue 
would require further clarification.

Finally, one can assume that the decomposition in Eq.~\eqref{eq:e1} also holds at the critical point~\cite{Wald20-1}. This gives 
\begin{equation}
 \bra{1}\mathbb{X}^{\rm (th)}\ket{1} \sim L, \quad
 \bra{1}\mathbb{X}^{(L)}\ket{1} \sim L, \quad
 \bra{1}\mathbb{P}^{\rm (th)}\ket{1} \sim \frac{\ln(L)}{L}, \quad
 \bra{1}\mathbb{P}^{(L)}\ket{1} \sim \frac{\ln(L)}{L}.
\end{equation}
This implies that the entanglement gap vanishes as $\delta \xi \sim 1/\sqrt{\ln(L)}$. 
Although this scaling is not correct, reflecting that Eq.~\eqref{eq:e1} does not hold
at criticality, 
it still captures the logarithmic character of the entanglement gap.

\section{Entanglement gap in 1D QSM with long-range interactions}
\label{sec:1d}
Recently, there has been increasing interest in quantum systems with long-range  
interactions~\cite{Defenu21,Def21}, also due to significant experimental advances~\cite{Zhang2017}. 
Since long-range interactions affect the structure of quantum 
correlations between subsystems, it is interesting to study entanglement witnesses in these systems.
Indeed, the study of entanglement in long-range systems has seen a significant surge of 
interest~\cite{Koff12,Raja13,vodola2016long,frerot2017entanglement,gong2017entanglement,mozaffar2017entanglement,maghrebi2017continuous,pappalardi2018scrambling,mozaffar2019entanglement,bentsen2022entanglement,ares2022symmetry}.

Arguably one of the paradigmatic systems is the long-range QSM at $T=0$ in one spatial 
dimension that we introduced
in section~\ref{sec:sm}. In terms of the long-range exponent $\alpha$ the 
interaction strength between 
two lattice sites behaves as $u_{nm}\sim |n-m|^{-(1+\alpha)}$. The parameter $\alpha$ satisfies 
$0\leq\alpha<2$ where $\alpha = 2$ would correspond to nearest-neighbor interaction and $\alpha=0$
is essentially an infinite range interaction. 
The zero temperature phase diagram is reported in Fig.~\ref{fig:pds} (right panel).  
 For $0<\alpha<2/3$ 
 the transition is of the mean-field type, whereas  for $2/3<\alpha<2$ the model shows a non-mean-field 
transition~\cite{Vojta96,Wald23}.

Crucially, despite the long-range nature of the model, in the ordered phase 
Eq.~\eqref{eq:e1} holds true. 
Thus, the study of the scaling of the entanglement gap proceeds as outlined 
in section~\ref{sec:order}. The 
susceptibilities $\chi^x$ and $\chi^t$ can be analyzed for $L\to\infty$ 
with regularization techniques involving the Mellin 
transform~\cite{contino-2002,Wald23} (details can be found in Ref.~\cite{Wald23}).
\begin{figure}[t]
 \centering
 \includegraphics[width = .32\textwidth]{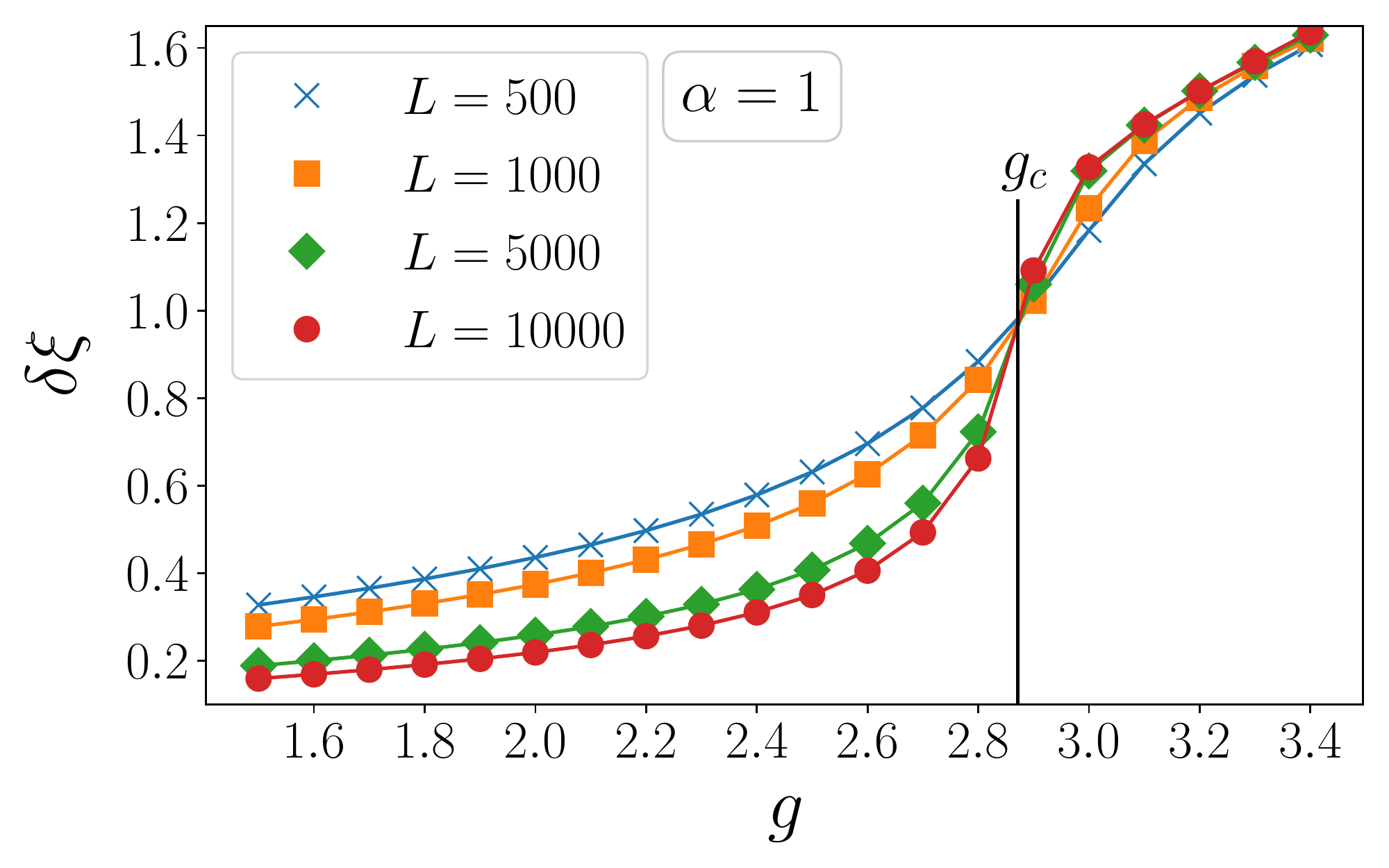}
 \includegraphics[width = .32\textwidth]{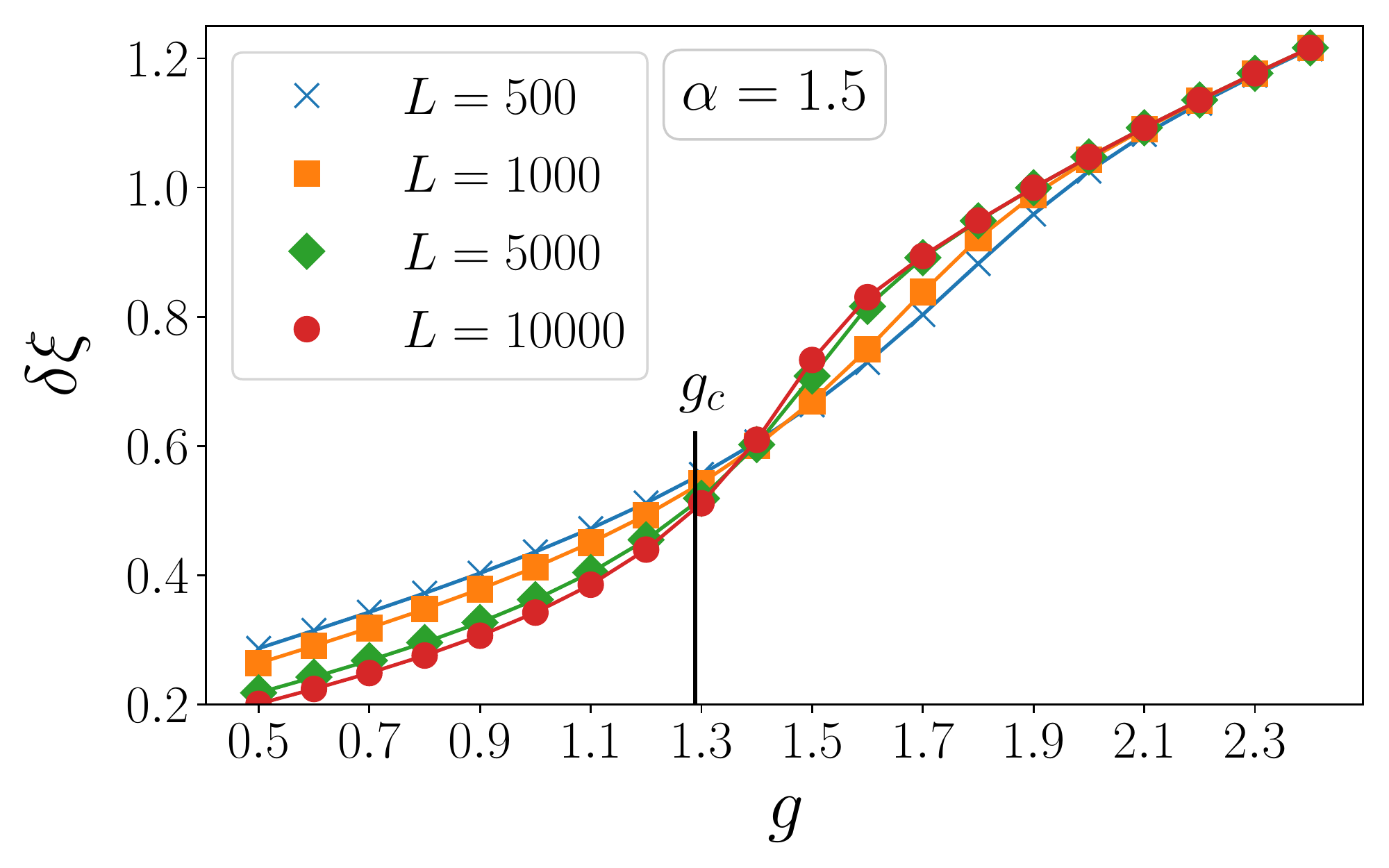}
 \includegraphics[width = .32\textwidth]{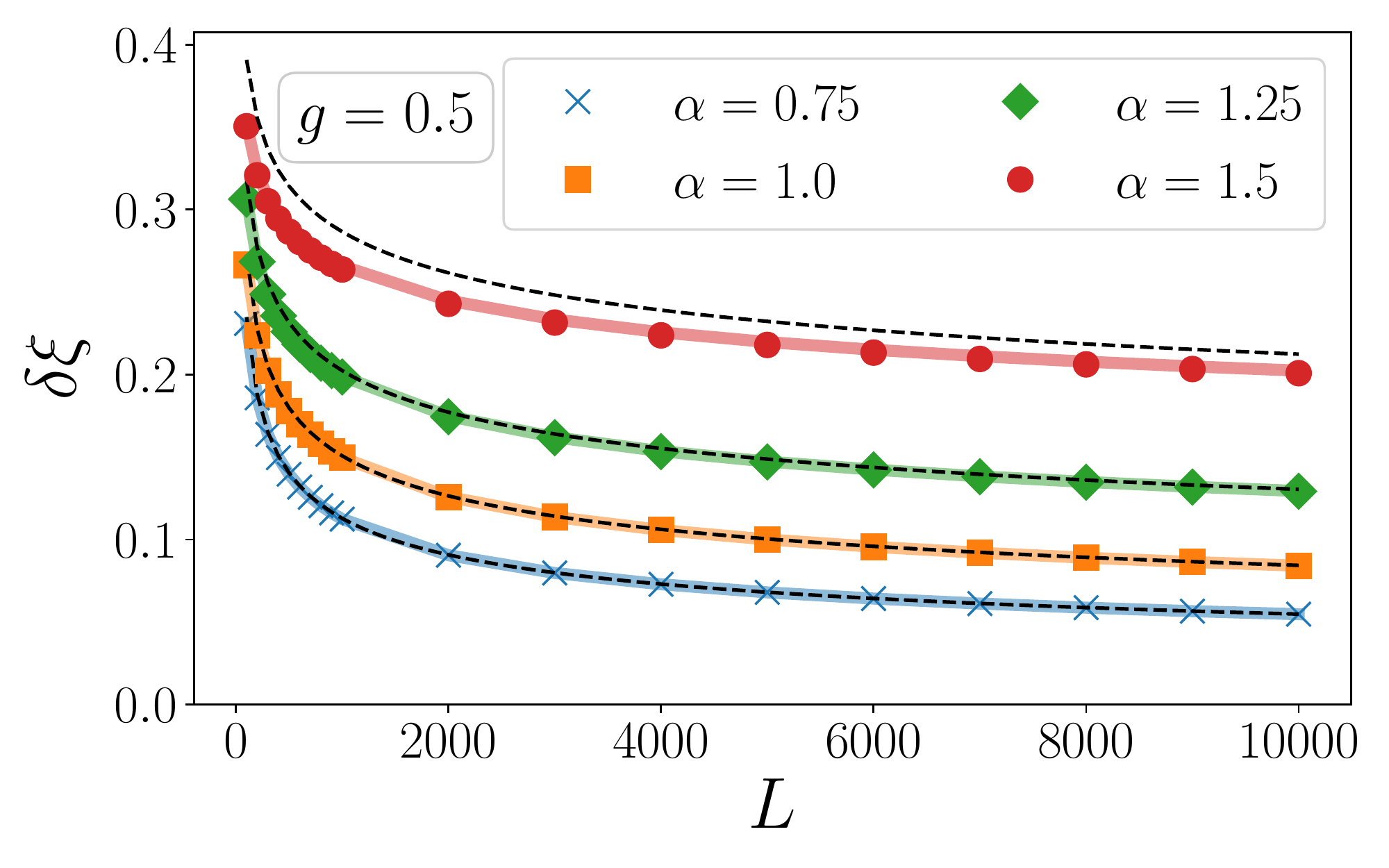}
 \caption{{\bf Entanglement gap in the long-range $1d$ QSM}: 
	 The left and center panel show the entanglement gap 
 for different values of the long-range exponent $\alpha$ 
 ($\alpha=1$ and $\alpha = 1.5$ respectively)
 across the phase transition. We observe a qualitative difference between the paramagnetic phase and 
 the ferromagnetic phase.
 The right panel shows the closure of the entanglement gap for different values of $\alpha$ in the 
 ferromagnetic phase. Symbols
 are numerical results from exact diagonalization and the continuous line corresponds to~Eq.~\eqref{eq:e1}. Black dashed lines correspond to the analytical prediction.
 }
 \label{fig:xi1d}
\end{figure}

In Fig.~\ref{fig:xi1d} we present in the left and center panel a numerical analysis of the 
entanglement gap across the phase diagram. We observe that $\delta\xi$ 
remains finite in the paramagnetic phase, whereas it shows a vanishing behavior in 
the ferromagnetic phase upon increasing $L$. 
As in the two dimensional QSM it is useful to decompose the correlation matrices into a thermodynamic
and a finite size part. Using Mellin techniques we obtain~\cite{Wald23} 
\begin{align}
 \bra{1}\mathbb{X}^{\rm (th)}\ket{1} \sim L, \quad
 \bra{1}\mathbb{X}^{(L)}\ket{1} \sim L^{\alpha/2}, \quad
 \bra{1}\mathbb{P}^{\rm (th)}\ket{1} \sim L^{-\alpha/2}, \quad
 \bra{1}\mathbb{P}^{(L)}\ket{1} \sim L^{-\alpha/2}.
\end{align}
Further details such as the precise prefactors, subleading contributions, the behavior at criticality
and the numerical benchmarks for all these results can be found in Ref.~\cite{Wald23}. These results allow 
us to deduce the entanglement gap using Eq.~\eqref{eq:e1}. 
We obtain 
\begin{align}
\label{eq:xi1d}
 \delta\xi \sim L^{-(1/2-\alpha/4)}.
\end{align}
In the right panel of Fig.~\ref{fig:xi1d} we show a comparison of exact diagonalization results 
and Eq.~\eqref{eq:xi1d}, finding perfect agreement.
Note that the entanglement gap decays algebraically with $L$, 
and multiplicative logarithmic corrections are absent. This differs 
strictly from the logarithmic behavior encountered in the previous section.

%##################################
\section{Summary and Conclusions}
\label{sec:conclusion}

We provided an overview of several results on the entanglement scaling in the QSM.
In particular, we investigated a variety of scenarios comprising 
$d=1,2,3$ dimensional quantum systems with long and short range interactions.

Precisely, we discussed the interplay of classical and 
quantum fluctuations at thermal transitions in the $3d$ QSM. We presented results for the 
entanglement entropy, the entanglement negativity and the entanglement spectrum in Sec.~\ref{sec:3d}. 
In particular, we mapped out the negativity across the 
whole phase diagram (see Fig.~\ref{fig:negativity}).
A more detailed analysis can be found in our work in Ref.~\cite{Wald20}. 
In Sec.~\ref{sec:2d} and Sec.~\ref{sec:order} we  
focussed on the quantum phase transition and on the ferromagnetic phase 
at zero temperature in two spatial dimensions. We discussed 
the behavior of the entanglement gap in the different quantum phases and at criticality. 
We showed that the entanglement gap is capable of detecting criticality in the QSM, 
although logarithmic corrections are present. 
We refer to Refs.~\cite{Wald20-1,alba2020entanglement} for  more 
details on the entanglement spectrum, the precise finite-size scaling, and for 
the study of the effect of corners in the entanglement spectrum. 
Finally, we reviewed entanglement properties of a one dimensional spherical quantum chain
with long-range interactions~\cite{Wald23}. Again, we considered the behavior of the entanglement 
gap across the zero temperature quantum phase transition for different long-range interactions. 
Remarkably, the QSM allows for a detailed analytical investigation of the entanglement 
gap in the ordered phase of the model. In particular, it is possible to understand how the 
entanglement gap is affected by the long-range nature of the interactions.

The spherical model - here in its quantum formulation - has again proven itself as a remarkably
useful tool to study collective phenomena in strongly interacting many-body systems. 
Clearly, the QSM will continue to serve as a reference system for future studies 
of entanglement-related quantities. For example it would be interesting to 
further investigate the 
influence of corners on the entanglement patterns. 
Furthermore, it would be enlightening to consider the QSM on quasi two-dimensional structures, 
such as ladders. It would
also be interesting to study the influence of disorder, or explore the full entanglement Hamiltonian
explicitly. Moreover, it  has been shown that the criticality in the QSM (in and out of equilibrium)
can be exploited as a resource in quantum metrology~\cite{Wald20-2}. It would be 
interesting to explore to which extend entanglement might affect
and support quantum metrology protocols. 
Furthermore, non-equilibrium and relaxational quantum dynamics has been 
extensively studied in the QSM in the past years~\cite{Wald16,Wald18,Timpa19,Wald21,Maraga15,Henk22}.
It would be interesting to derive the spreading of entanglement in such scenarios in the QSM, possibly
exploiting the results from Refs.~\cite{Chan13,barbier2019pre,barbier2022generalized,Henk22}. 
An intriguing idea is to study the entanglement using
the Kibble-Zurek dynamics~\cite{Scopa18,ROSSINI20211}.

\section*{Acknowledgement}
It is our pleasure to dedicate this work to our friend and colleague Malte Henkel on the occasion of
his $60^{\rm th}$ birthday. 
SW would like to further thank Malte for years of excellent and interesting 
collaborations, exchanges and guidance.

% \section*{Data Availability Statement}
% No Data associated in the manuscript.

%##################################
\section*{References}
\bibliographystyle{iopart-num.bst}
\bibliography{bibliography}

\appendix

\end{document}